\documentclass[aps,prd,superscriptaddress,notitlepage,showpacs,showkeys,10pt]{revtex4-1}

\usepackage{graphicx}
\usepackage{amsmath}
\usepackage{amssymb}
\usepackage{color}
\usepackage{bm}

\definecolor{purple}{rgb}{0.8,0,0.6}
\definecolor{darkgreen}{rgb}{0.00,0.6,0.00}
\begin{document}

\title{Pseudomagnetic lens as a valley and chirality splitter in Dirac and Weyl materials}

\author{E.~V.~Gorbar}
%\email{gorbar@bitp.kiev.ua}
\affiliation{Department of Physics, Taras Shevchenko National Kiev University, Kiev, 03680, Ukraine}
\affiliation{Bogolyubov Institute for Theoretical Physics, Kiev, 03680, Ukraine}

\author{V.~A.~Miransky}
%\email{vmiransk@uwo.ca}
\affiliation{Department of Applied Mathematics, Western University, London, Ontario N6A 5B7, Canada}
\affiliation{Department of Physics and Astronomy, Western University, London, Ontario N6A 3K7, Canada}

\author{I.~A.~Shovkovy}
%\email{igor.shovkovy@asu.edu}
\affiliation{College of Integrative Sciences and Arts, Arizona State University, Mesa, Arizona 85212, USA}
\affiliation{Department of Physics, Arizona State University, Tempe, Arizona 85287, USA}

\author{P.~O.~Sukhachov}
%\email{psukhach@uwo.ca}
\affiliation{Department of Applied Mathematics, Western University, London, Ontario N6A 5B7, Canada}

\begin{abstract}
It is proposed that strain-induced pseudomagnetic fields in Dirac and Weyl materials could be used as
valley and chirality sensitive lenses for beams of Weyl quasiparticles. The study of the (pseudo-)magnetic
lenses is performed by using the eikonal approximation for describing the Weyl quasiparticles propagation
in magnetic and strain-induced pseudomagnetic fields. Analytical expressions for the locations of
the principal foci and the focal lengths are obtained in the paraxial approximation in the models
with uniform as well as nonuniform effective magnetic fields inside the lens. The results show
that the left- and right-handed quasiparticles can be focused at different spatial locations when
both magnetic and pseudomagnetic fields are applied. It is suggested that the use of magnetic
and pseudomagnetic lenses could open new ways of producing and manipulating beams of
chiral Weyl quasiparticles.
\end{abstract}


%\pacs{03.65.Sq}
%71.45.-dCollective effects
%71.45.GmExchange, correlation, dielectric and magnetic response functions, plasmons
%03.65.SqSemiclassical theories and applications
%72.30.+qHigh-frequency effects; plasma effects

\maketitle

%\tableofcontents
%\newpage

{\it Introduction.} ---
The recent experimental discovery of Dirac (e.g., $\mathrm{Na_3Bi}$ and $\mathrm{Cd_3As_2}$
\cite{Borisenko,Neupane,Liu}) and Weyl (e.g., $\mathrm{TaAs}$, $\mathrm{TaP}$, $\mathrm{NbAs}$,
$\mathrm{NbP}$, $\mathrm{Mo_xW_{1-x}Te}$, and $\mathrm{YbMnBi_2}$
\cite{Tong,Bian,Qian,Long,Belopolski,Cava}) materials proved a conceptual possibility of
condensed-matter systems whose low-energy quasiparticles are massless Dirac or Weyl
fermions (for reviews, see Refs.~\cite{Turner,Vafek:2013mpa,Burkov:2015}). These
discoveries opened a new chapter in studies of the effects associated with the quantum
anomalies, e.g., the chiral anomaly \cite{ABJ} in parallel electric and magnetic fields, by
using simple table-top experiments, rather than accelerator techniques of high-energy
physics. The Dirac and Weyl materials not only mimic the properties of truly relativistic
matter but also allow for the realization of novel quantum phenomena that cannot exist
in high-energy physics. In particular, a very intriguing example of such phenomena is
the quasiparticle response to background pseudoelectromagnetic (axial) fields. Unlike
the ordinary electromagnetic fields $\mathbf{E}$ and $\mathbf{B}$, their pseudoelectromagnetic
counterparts $\mathbf{E}_5$ and $\mathbf{B}_5$ couple to the left- and right-handed
quasiparticles with opposite signs. It was shown in
Refs.~\cite{Zubkov:2015,Cortijo:2016yph,Cortijo:2016wnf,Grushin-Vishwanath:2016,Pikulin:2016,Liu-Pikulin:2016}
that similarly to graphene, the physical origin of the pseudomagnetic fields is related
to deformations in Dirac and Weyl materials, which cause unequal modifications of
hopping parameters in strained crystals and can be considered as effective axial gauge fields.
The characteristic strengths of the pseudomagnetic fields in Dirac and Weyl semimetals
are much smaller than in graphene and range from about $B_5\approx0.3~\mbox{T}$,
when a static torsion is applied to a nanowire of Cd$_3$As$_2$ \cite{Pikulin:2016}, to
approximately $B_5\approx15~\mbox{T}$, when a thin film of Cd$_3$As$_2$ is bent
\cite{Liu-Pikulin:2016}.

While a static magnetic field cannot change the kinetic energy of charged particles,
it affects the direction of their motion due to the Lorentz force. This property is widely
utilized in physics and technology. In particular, magnetic fields can be employed to create magnetic lenses
for deflecting and focusing beams of charged particles \cite{Busch,Knoll}. Such lenses
are employed, for example, in cathode ray tubes and electron microscopes \cite{Rose,Reizer}.

\begin{figure}[t]
\begin{center}
\includegraphics[width=0.5\textwidth]{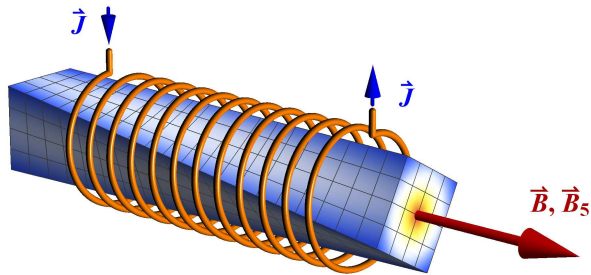}
\end{center}
\caption{A schematic illustration of the experimental setup which allows for the magnetic and pseudomagnetic
lensing of quasiparticles in Weyl materials. While the usual magnetic field $\mathbf{B}$ is produced by an
electric current in the solenoid, the pseudomagnetic field $\mathbf{B}_5$ is created by twisting
the crystal of a Weyl material.}
\label{fig:illustration}
\end{figure}

In this study we suggest that strain-induced pseudomagnetic fields can be used for creating
pseudomagnetic lenses which deflect and focus beams of Weyl quasiparticles depending on their
chirality. Such a dependence can be used to spatially separate the charged quasiparticles of
different chirality. A general experimental setup that allows the maximum control of chiral beams
is given by a combination of the magnetic and pseudomagnetic lenses as shown schematically
in Fig.~\ref{fig:illustration}. The system consists of a Weyl crystal (wire) placed
inside a solenoid. The magnetic and pseudomagnetic fields are directed along the $+z$
axis and are present in the region $0<z<L$. The magnetic field is generated by an electric
current in the solenoid, and the pseudomagnetic one is produced by the torsion of the crystal.
When the system is a part of a circuit, an input electric current will induce an unpolarized stream
of left- and right-handed chiral quasiparticles inside the semimetal. After passing
through the lens region $0<z<L$, the quasiparticles of opposite chiralities will split
and converge at different spatial locations, providing a steady state with a localized
chiral asymmetry.

Since the characteristic scales of spatial variations of the background magnetic and
pseudomagnetic fields are much larger than the de Broglie wavelengths of Weyl
quasiparticles, one can use the methods of ray optics (namely the eikonal approximation)
for the description of their motion (see, e.g., Ref.~\cite{Landau:t2}). Because of a
nontrivial topology of chiral quasiparticles, however, one should pay special attention
to the Berry curvature \cite{Berry:1984} effects on the corresponding chiral beams.

%%%%%%%%%%%%%%%%%%
%%%%%%%%%%%%%%%%%%

{\it Eikonal approximation for Weyl quasiparticles.} ---
Let us start with the formulation of the eikonal approximation for the motion of Weyl quasiparticles
in the presence of both magnetic and pseudomagnetic fields. In view of the nontrivial topological
properties of Weyl fermions \cite{Niu,Xiao,Duval,Gao:2015}, the semiclassical equations of
motion should take into account the Berry curvature effects \cite{Berry:1984,Xiao:2009rm}.
In the framework of the chiral kinetic theory \cite{Son:2012wh,Stephanov,Son:2012zy},
it is straightforward to include corrections linear in the background field
$\mathbf{B}_{\lambda} = \mathbf{B}+\lambda \mathbf{B}_{5}$, where $\mathbf{B}$ and
$\mathbf{B}_{5}$ denote the ordinary magnetic and pseudomagnetic fields, respectively,
and $\lambda$ is chirality of the left- ($\lambda=-$) and right-handed ($\lambda=+$)
quasiparticles. In the eikonal approximation \cite{Landau:t2}, however, we
need the dispersion relations for the Weyl quasiparticles
valid to the second order in $\mathbf{B}_{\lambda}$. While the
corresponding expression for the general band structure was derived in
Ref.~\cite{Gao:2015}, its explicit form for Weyl quasiparticles was found
by the present authors in Ref.~\cite{Gorbar:2017cwv}. For the quasiparticles
of positive energy (electrons), the corresponding relation reads
\begin{eqnarray}
\varepsilon\equiv v_Fp
-\lambda\frac{e\hbar v_F }{2cp^2} (\mathbf{B}_{\lambda}\cdot\mathbf{p}) + \frac{e^2 \hbar^2 v_F}{16c^2p^3}
\left(2\mathbf{B}_{\lambda}^2-\frac{(\mathbf{B}_{\lambda}\cdot\mathbf{p})^2}{p^2}\right),
\label{second-energy-2}
\end{eqnarray}
where $v_F$ is the Fermi velocity, $c$ is the speed of light, $\mathbf{p}$ is the momentum of
quasiparticles, and $e<0$ is the electron charge. Note that the second and third terms in
Eq.~(\ref{second-energy-2}) describe corrections due to the Berry curvature. In essence,
these terms describe the interaction of the spin magnetic moment of the quasiparticles with
the effective magnetic field. Further, we use the notation $\varepsilon$ without the subscript
$\lambda$ because the quasiparticles of opposite chiralities have the same Fermi energy.

The orbital part of the quasiparticle interaction with the magnetic field is captured in the
standard eikonal approximation \cite{Rose,Landau:t2}. (See Sec.~I of the Supplemental
Material %\cite{pseudomagnetic-lens-supplemental} 
for the key details of the
eikonal approximation.) For charged quasiparticles in the
effective magnetic field $\mathbf{B}_{\lambda}$ close to the optical axis, we can write down the abbreviated action $S_0\equiv S_0(\mathbf{r})$
for Weyl quasiparticles in the following form:
\begin{equation}
S_0\approx \frac{\varepsilon}{v_F}\left( C z +\frac{r_{\perp}^2}{2}A(z) +O(r_{\perp}^4) \right),
\label{lens-eikonal-S0}
\end{equation}
where $r_{\perp}=\sqrt{x^2+y^2}$ measures the distance from the optical axis. (Note that the
abbreviated action $S_0$ is related to the full action via $S=-\varepsilon t +S_0$, where
$\varepsilon$ denotes the quasiparticle energy and $t$ is time.) By solving the eikonal
equation in the weak-field limit, we obtain the explicit expression for the constant $C$
\begin{equation}
C \simeq 1 + \lambda \frac{B_{\lambda}}{B^{*}}
- \frac{5}{4}\left(\frac{B_{\lambda}}{B^{*}}\right)^2 +O\left(\frac{B_{\lambda}^3}{(B^{*})^3}\right) ,
\label{lens-eikonal-epsilon-r-0-sol-main}
\end{equation}
and the equation for the function $A(z)$
\begin{equation}
a_1 A^{\prime}(z) +(A(z))^2 +a_2^2 =0.
\label{lens-eikonal-epsilon-r-1-comp}
\end{equation}
Here we introduced the following notations:
\begin{eqnarray}
\label{lens-eikonal-epsilon-r-1-a1}
a_1&\simeq& 1+\frac{5}{4} \left(\frac{B_{\lambda}}{B^{*}}\right)^2, \\
a_2^2&\simeq& \frac{ eB_{\lambda}^2}{2c \hbar B^{*}}  \left(1-2\lambda \frac{B_{\lambda}}{B^{*}} \right),
\label{lens-eikonal-epsilon-r-1-a2}
\end{eqnarray}
as well as a reference value of the magnetic field, $B^{*} = 2c \varepsilon^2/(e\hbar v_F^2)$,
which is associated with the quasiparticle energy scale $\varepsilon$. The latter is defined
so that the corresponding magnetic length $l^{*}_{B}\equiv \sqrt{c\hbar/(eB^{*})}$ is
comparable to the de Broglie wavelength of the Weyl quasiparticles
$l_{\varepsilon}\equiv\hbar v_F /\varepsilon=\sqrt{2}\,l^{*}_{B}$. As is easy to check,
the subleading terms in powers of $B_{\lambda}/B^{*}$ in
Eqs.~(\ref{lens-eikonal-epsilon-r-0-sol-main}), (\ref{lens-eikonal-epsilon-r-1-a1}), and
(\ref{lens-eikonal-epsilon-r-1-a2}) originate from the Berry curvature corrections in
the dispersion relation (\ref{second-energy-2}). Their validity, therefore, is similarly
restricted to the case of sufficiently weak effective field, i.e., $B_{\lambda}\ll B^{*}$.

{\it Magnetic and pseudomagnetic lenses.} ---
Let us begin the analysis of the quasiparticle motion with the simplest case of uniform magnetic
and pseudomagnetic fields, $B_{\lambda}=const$. Solving Eq.~(\ref{lens-eikonal-epsilon-r-1-comp})
in the three different regions, i.e., $z<0$, $0<z<L$, $z>L$, and matching the abbreviated
action at the boundaries (see Sec.~II of the Supplemental Material %\cite{pseudomagnetic-lens-supplemental}
for the details of solving the lens equation in a uniform field), we obtain the following
lens equation relating the coordinates of the quasiparticle source $z_1$ and its image $z_2$:
\begin{equation}
(z_1 + g_{\lambda})(z_2 - h_{\lambda})=-f^2_{\lambda}.
\label{lens-eikonal-lens-eq-main}
\end{equation}
Indeed, one can easily see that when the source is placed at the left focal point, i.e.,
$z_1 \to -g_{\lambda}$, the position of the image $z_2$ goes to infinity. Similarly,
when $z_1 \to -\infty$, the location of the image is near the right focal point, i.e., $z_2 \to h_{\lambda}$.
Therefore, $z=-g_{\lambda}$ and $z=h_{\lambda}$ are the locations of the principal foci,
and $f_{\lambda}$ is the principal focal length. In the case under consideration, we find
that $h_{\lambda} = L+g_{\lambda}$ and
\begin{eqnarray}
\label{lens-eikonal-lens-g-app-2}
g_{\lambda} &\simeq &\frac{ l_{\varepsilon} B^{*} }{B_{\lambda} \left(1-\lambda B_{\lambda}/B^{*}\right)}
\cot{\left(\frac{L B_{\lambda}\left(1-\lambda B_{\lambda}/B^{*}\right)}{l_{\varepsilon} B^{*}}\right)},\\
\label{lens-eikonal-lens-f-app-2}
f_{\lambda} &\simeq & \frac{ l_{\varepsilon} B^{*}}{B_{\lambda} \left(1-\lambda B_{\lambda}/B^{*}\right)
\sin {\left(\frac{L B_{\lambda}\left(1-\lambda B_{\lambda}/B^{*}\right)}{l_{\varepsilon} B^{*}}\right)} }.
\end{eqnarray}
These analytical expressions are the key characteristics of the (pseudo-)magnetic lens and
are the main results of this article. When the paraxial approximation is justified, these results
should be valid for arbitrary Weyl and Dirac materials.

According to Eq.~(\ref{lens-eikonal-lens-f-app-2}), the focal lengths
for the quasiparticles of opposite chiralities can be different. In fact, this remains
true even in the limit of the vanishing pseudomagnetic field (i.e., $B_5=0$ but $B\neq 0$). In such
a case, a relatively small difference between the focal lengths $f_{+}$ and $f_{-}$ is connected
with the Berry curvature effects quantified by the second term in the parentheses in Eq.~(\ref{lens-eikonal-epsilon-r-1-a2}).
This is in contrast to the case of the vanishing magnetic field ($B=0$ but  $B_5\neq 0$),
when the focal lengths for the quasiparticles of opposite chiralities are exactly the same.
The latter should not be surprising after noting that the Berry curvature effects, which are
proportional to $\lambda B_{\lambda}/B^{*}$, are identical for the left- and right-handed
quasiparticles when $B=0$. However, it is important to mention that comparing to the
case of nonzero magnetic and pseudomagnetic fields, where the chirality splitting shows up already
at the leading linear order in the fields, the Berry curvature effects are quadratic in the fields.
Therefore, the corresponding splitting is much smaller than that caused by the combination
of magnetic and pseudomagnetic fields.

\begin{figure}
\begin{center}
\includegraphics[width=0.45\textwidth]{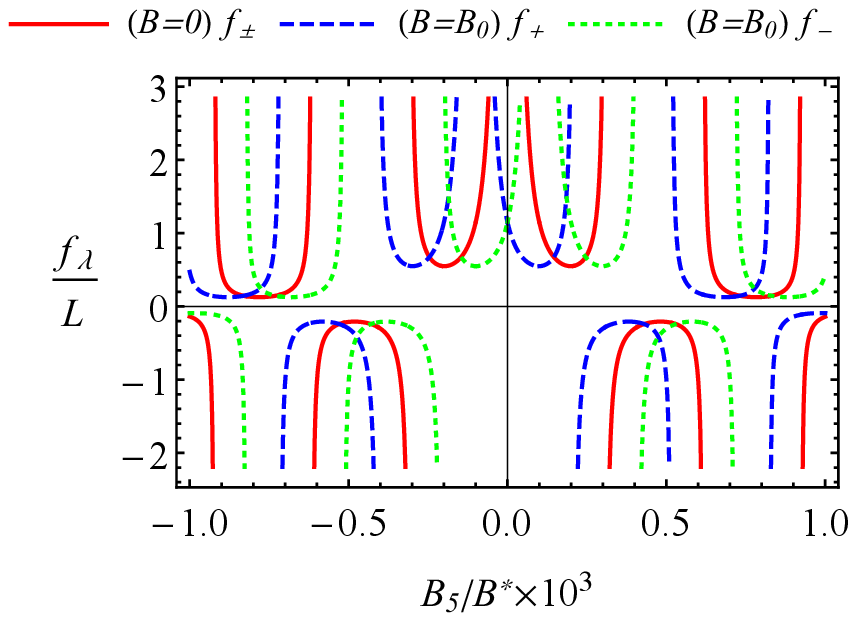}\hfill
\includegraphics[width=0.45\textwidth]{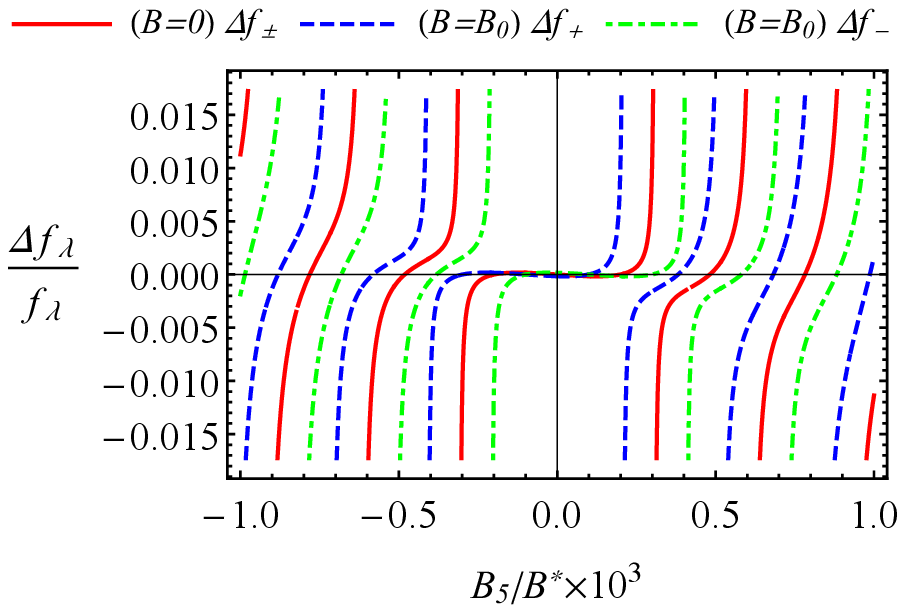}
\end{center}
\caption{Left panel:
The focal length $f_{\lambda}$ for the quasiparticles of given chirality $\lambda$ as a function of the
pseudomagnetic field. Right panel:  The relative difference between the focal lengths with and without
the Berry curvature effects $\Delta f_{\lambda}/f_{\lambda}$ as a function of the pseudomagnetic field.
While the red solid lines correspond to quasiparticles of both chiralities at $B=0$, the blue dashed and
green dotted lines represent the right- and left-handed quasiparticles at $B=10^{-4} B^{*}$, respectively.
We set $\varepsilon=100~\mbox{meV}$ and $L=10^{-2}~\mbox{cm}$.}
\label{fig:focal-f-B5-B}
\end{figure}

In order to get a quantitative estimate for the principal focal length, we will use the numerical
value of the Fermi velocity for $\mathrm{Cd_3As_2}$ \cite{Neupane}, $v_F\approx9.8~\mbox{eV\AA}$.
In this case, the characteristic magnetic field is $|B^{*}|\approx 14~\mbox{T}$, assuming
$\varepsilon=100~\mbox{meV}$. The dependence of the focal lengths $f_+$ and $f_-$ on the
pseudomagnetic field strength $B_5$ is shown in the left panel of Fig.~\ref{fig:focal-f-B5-B} for
$B=0$ and $B=10^{-4}B^{*}$. As expected, in the presence of both magnetic and
pseudomagnetic fields, the focal lengths are different for the quasiparticles of opposite chirality.

As we see from the analytical expression (\ref{lens-eikonal-lens-f-app-2}), as well as from
the left panel of Fig.~\ref{fig:focal-f-B5-B}, the dependence of the focal length
on the (pseudo-)magnetic field is quasiperiodic. Also, the focal length is formally divergent at the
following discrete values of the background field: $B_{\lambda}/B^{*}\approx l_{\varepsilon}\pi n/L$,
where $n=0,\pm1,\pm2,\ldots$. The mathematical reason for these divergencies is clear from
Eq.~(\ref{lens-eikonal-lens-f-app-2}), which has the sine function in the denominator. In the
vicinity of divergencies, the paraxial approximation breaks down. This follows from the
fact that the expansion in powers of $r_\perp$ in Eq.~(\ref{lens-eikonal-S0}) becomes unreliable
when $A(z)$ is too large. Thus, the geometrical optics approach fails and one needs
to use exact, rather than approximate solutions to the wave equation.

%%%%%%%%%%%%%%%%%%
%%%%%%%%%%%%%%%%%%

In order to illuminate the effects of the Berry curvature, it is instructive to compare the focal length in
Eq.~(\ref{lens-eikonal-lens-f-app-2}) with its counterpart when such effects are neglected, i.e.,
\begin{eqnarray}
f_{\lambda}^{(0)} = \frac{l_{\varepsilon} B^{*}}{B_\lambda  \sin{\left(\frac{L B_\lambda}{l_{\varepsilon}B^{*}}\right)}}.
\label{lens-eikonal-lens-f-app-Landau}
\end{eqnarray}
The corresponding relative difference $\Delta f_{\lambda}/f_{\lambda} \equiv (f_{\lambda}^{(0)}-f_{\lambda})/f_{\lambda}$
is plotted in the right panel of Fig.~\ref{fig:focal-f-B5-B} for $B=0$ and $B=10^{-4}B^{*}$.
As one can see, away from the divergencies, the effects of the Berry curvature quantified by
$\Delta f_{\lambda}$ are much smaller than the focal lengths. This fact is not surprising because
the effects of the Berry curvature in Weyl materials are subleading compared to that of the
combination of magnetic and pseudomagnetic fields. However, the corresponding effects
may become noticeable at sufficiently large magnetic and pseudomagnetic fields or near
the divergencies, where the Berry curvature corrections may lead to a noticeable spatial
separation of the beams of chiral quasiparticles even by the ordinary magnetic field alone.

\begin{figure}
\begin{center}
\includegraphics[width=0.5\textwidth]{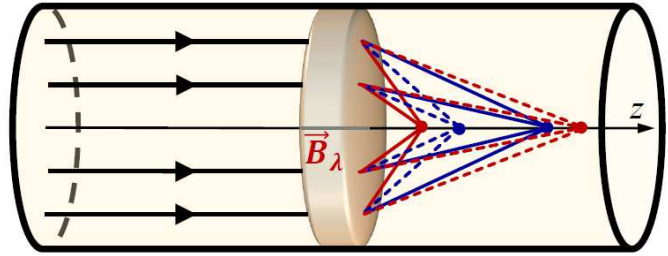}
\end{center}
\caption{A schematic illustration of the spatial separation of chiral quasiparticles inside a Dirac
semimetal crystal that is a  $\mathbb{Z}_2$ Weyl semimetal. While the red lines correspond to
the right-handed quasiparticles, the blue lines represent the left-handed ones. The solid and
dashed lines correspond to the different pairs of Weyl nodes.}
\label{fig:focal-f-Dirac}
\end{figure}

Just like in Weyl materials, a combination of the magnetic and pseudomagnetic lenses can
be used for splitting the beams of chiral quasiparticles in Dirac materials. Naively, by taking
into account the topological triviality of a Dirac point, one may suggest that the pseudomagnetic
fields are absent and the spatial separation of the quasiparticles of different
chiralities in magnetic and pseudomagnetic fields is impossible. However, some Dirac
semimetals, e.g., $\mathrm{A_3Bi}$ ($\mathrm{A}=\mathrm{Na}$, $\mathrm{K}$,
$\mathrm{Rb}$) as well as certain phases of $\mathrm{Cd_3As_2}$, are, in fact,
hidden $\mathbb{Z}_2$ Weyl semimetals \cite{Gorbar:2014sja}. Their Dirac points come
from two pairs of superimposed Weyl nodes separated in the momentum space. Since the
strain-induced pseudomagnetic field $\mathbf{B}_{5}$ is determined by the separation
vector \cite{Cortijo:2016yph,Pikulin:2016,Liu-Pikulin:2016}, this field will be the same in
magnitude but opposite in direction for the corresponding two pairs of Weyl nodes. Therefore,
if the Berry curvature effects were neglected as given by Eq.~(\ref{lens-eikonal-lens-f-app-Landau}),
the focal length of the right-handed (left-handed) quasiparticles from the first pair of Weyl
nodes would coincide exactly with the focal length of the left-handed (right-handed)
quasiparticles of the second pair of Weyl nodes. As a result, a generic beam of
quasiparticles would split in two nonchiral beams after passing through the lens.
In other words, there would be only valley separation with no spatial separation
of the chirality. However, the Berry curvature qualitatively changes the situation
and, according to Eq.~(\ref{lens-eikonal-lens-f-app-2}), all focal lengths become
different now. This is schematically illustrated in Fig.~\ref{fig:focal-f-Dirac}, where
the beams of quasiparticles in Dirac semimetals under consideration are split into
four chiral beams after passing through the lens. In summary, while the combination
of magnetic and pseudomagnetic fields alone allows only for a valley separation in
the Dirac materials, the inclusion of the Berry curvature effects is important in order
to achieve the complete splitting of the beams. This is in contrast to the case of Weyl
materials with two Weyl nodes where the Berry curvature provides only second-order
corrections to the splitting induced by the (pseudo-)magnetic fields at the linear order.

%%%%%%%%%%%%%%%%%%
%%%%%%%%%%%%%%%%%%

{\it Lenses with a nonuniform background field.} ---
It could be argued that the model situation with a constant \mbox{(pseudo-)magnetic} field
$\mathbf{B}_\lambda$ considered above may not be very realistic. Indeed, for real solenoids
and torsion-induced strains, there are always fringing fields at the ends. In order to get a
better insight into the underlying physics, it is instructive to consider the case of a nonuniform
(pseudo-)magnetic field with the following spatial profile:
\begin{equation}
\tilde{\mathbf{B}}_{\lambda}(z) = \theta(z)\theta\left(L-z\right) \frac{\mathbf{B}_{\lambda}}{1+\left(z-L/2\right)^2/\xi^2}.
\label{inhom-B-profile}
\end{equation}
Such a configuration mimics well the fringing fields at the ends of the (pseudo-)magnetic lens
and, at the same time, allows one to obtain an analytical solution in the paraxial approximation.
Because of the product of unit step functions $\theta(z)\theta\left(L-z\right)$ in Eq.~(\ref{inhom-B-profile}),
the effective field $\tilde{\mathbf{B}}_{\lambda}(z)$ is still effectively confined in the region $0<z<L$.

By limiting ourselves to the case of sufficiently weak (pseudo-)magnetic fields, we can neglect
the effects of the Berry curvature in the eikonal approximation. This is supported by our findings
presented in the right panel of Fig.~\ref{fig:focal-f-B5-B}, showing that the main contribution to
the spatial separation of Weyl quasiparticles with different chirality comes from the leading
linear order in the (pseudo-)magnetic field. Such an approximation also allows us to obtain
explicit analytical results. When the quadratic corrections due to the Berry curvature are
non-negligible, the corresponding analysis becomes much more involved but can be
performed using numerical methods.

As is easy to check in the case of a weak inhomogeneity, i.e., $\xi\geq L$, the focal length
$\tilde{f}_{\lambda}$ becomes almost indistinguishable from that in the uniform field shown
in the left panel of Fig.~\ref{fig:focal-f-B5-B}. Of course, this is expected for a weakly varying
(pseudo-)magnetic field. On the other hand, when the (pseudo-)magnetic field is sufficiently
nonuniform, i.e., $\xi\lesssim L$, the dependence of $\tilde{f}_{\lambda}$ on the
(pseudo-)magnetic field is different. (See Sec.~III of the Supplemental Material %\cite{pseudomagnetic-lens-supplemental}
for the details of solving the lens equation in a nonuniform field.)
We note, however, that one can still use
Eq.~(\ref{lens-eikonal-lens-f-app-Landau}) for $\xi\lesssim L$, but with the following
replacement of the lens size:
\begin{eqnarray}
L\to L_{\rm eff} = \frac{1}{B_{\lambda}^2} \int_{-\infty}^{\infty}dz \left(\tilde{B}_{\lambda}(z)\right)^2 = \frac{\pi \xi}{2},
\label{inhom-B-Leff}
\end{eqnarray}
where we used the explicit form of $\tilde{B}_{\lambda}(z)$ in Eq.~(\ref{inhom-B-profile})
in order to perform the integration. Thus, the nonuniform (pseudo-)magnetic fields do not
invalidate the principal possibility of pseudomagnetic lensing, albeit they may lead to some
technical challenges in its experimental realization.

{\it Conclusions.} ---
In this study we investigated the conceptual possibility of pseudomagnetic lenses that can be used to
focus the beams of chiral quasiparticles in Dirac and Weyl materials. We found that the maximum
flexibility in controlling the beams of Weyl quasiparticles is achieved in the presence of both magnetic
and pseudomagnetic fields. This allows one to achieve different magnitudes of the effective magnetic
fields exerted on the left- and right-handed quasiparticles, which provides enough control to manipulate
the focal lengths $f_{\lambda}$ independently for each chirality and valley. Moreover, in view of the
nontrivial topology of Weyl fermions, the corresponding eikonal equation is affected
by the Berry curvature leading to a splitting of the beams of chiral quasiparticles even by the ordinary
magnetic field alone, albeit with a much smaller amplitude. Its effects are, however, crucial for the
chirality separation in the Dirac semimetals, which comes on top of the valley separation induced by
(pseudo-)magnetic fields. A similar generic separation is also expected in multipair Weyl materials.

It is intriguing to suggest that the pseudomagnetic lenses could allow for new and powerful ways
of controlling and manipulating the chiral beams of quasiparticles inside Dirac and Weyl materials.
For example, they may open an experimental possibility to control the spatial distributions of the
electric current densities due to Weyl quasiparticles depending both on their valley and chirality
that may be detected via local probes. By focusing the beams of the left- and right-handed
quasiparticles in different spatial regions, one could achieve a ``chiral distillation" and/or steady
states of matter with a nonzero chiral asymmetry. We hope that findings of this study could lead
to new applications that utilize the beam splitters of charged particles of given valley and chirality.

\begin{acknowledgments}
The work of E.V.G. was partially supported by the Program of Fundamental Research of the Physics and
Astronomy Division of the National Academy of Sciences of Ukraine. The work of V.A.M. and P.O.S. was
supported by the Natural Sciences and Engineering Research Council of Canada. The work of I.A.S. was
supported by the U.S. National Science Foundation under Grant No.~PHY-1404232.
\end{acknowledgments}

%Supplemental materials in the main text
%%%%%%%%%% Merge with supplemental materials %%%%%%%%%%
\pagebreak
%\clearpage
\widetext
\begin{center}
\textbf{\large Supplemental Material: Pseudomagnetic lens as a valley and chirality splitter in Dirac and Weyl materials}
\end{center}
%%%%%%%%%% Merge with supplemental materials %%%%%%%%%%
%%%%%%%%%% Prefix a "S" to all equations, figures, tables and reset the counter %%%%%%%%%%
\setcounter{equation}{0}
\setcounter{figure}{0}
\setcounter{page}{1}
\makeatletter
\renewcommand{\theequation}{S\arabic{equation}}
\renewcommand{\thefigure}{S\arabic{figure}}
%\renewcommand{\bibnumfmt}[1]{[S#1]}
%\renewcommand{\citenumfont}[1]{S#1}
%%%%%%%%%% Prefix a "S" to all equations, figures, tables and reset the counter %%%%%%%%%%

\section{Key details of the eikonal approximation}

In this section, using the standard eikonal approximation for charged particles [S1],
we derive the equation for the abbreviated action $S_0\equiv S_0(\mathbf{r})$
of Weyl quasiparticles in the effective magnetic fields. We begin by making the following replacement
in the dispersion relation (1) in the main text:
\begin{equation}
\mathbf{p} \to \bm{\nabla} S_0 -\frac{e}{c}\mathbf{A}_{\lambda},
\label{lens-eikonal-p-replace}
\end{equation}
where $\mathbf{A}_{\lambda}=\left(-yB_{\lambda}/2, xB_{\lambda}/2,0\right)$ is an effective vector potential that describes the background field
$\mathbf{B}_{\lambda}$ in the $+z$ direction. Then, we obtain the
following eikonal equation for the abbreviated action $S_0$ of Weyl quasiparticles with energy $\varepsilon$:
\begin{eqnarray}
\varepsilon &=& v_F \sqrt{(\bm{\nabla}S_0)^2+\frac{e^2}{4c^2}r_{\perp}^2B_{\lambda}^2} - \frac{\lambda e \hbar v_F B_{\lambda}}{2c}\frac{\nabla_zS_0}{(\bm{\nabla}S_0)^2+\frac{e^2}{4c^2}r_{\perp}^2B_{\lambda}^2}
\nonumber\\
&+& \frac{e^2 \hbar^2v_FB_{\lambda}^2}{8c^2} \frac{1}{\left[(\bm{\nabla}S_0)^2+\frac{e^2}{4c^2}r_{\perp}^2B_{\lambda}^2\right]^{3/2}}
-\frac{e^2 \hbar^2v_FB_{\lambda}^2}{16c^2} \frac{(\nabla_zS_0)^2}{\left[(\bm{\nabla}S_0)^2+\frac{e^2}{4c^2}r_{\perp}^2B_{\lambda}^2\right]^{5/2}},
\label{lens-eikonal-epsilon}
\end{eqnarray}
where, in view of the symmetry in the problem, we assumed that $S_0(\mathbf{r})$ depends on
$z$ and $r_{\perp}=\sqrt{x^2+y^2}$, and used
\begin{equation}
\left(\bm{\nabla}S_0-\frac{e}{c}\mathbf{A}_{\lambda}\right)^2 = (\bm{\nabla}S_0)^2 +\frac{e^2}{4c^2} r_{\perp}^2B_{\lambda}^2.
\label{lens-eikonal-eq-S0}
\end{equation}
In order to obtain an analytical solution to Eq.~(\ref{lens-eikonal-epsilon}), we will use the paraxial approximation. In other words, we will assume
that $r_{\perp}$ is small and expand the solution in powers of $r_{\perp}$. Clearly, this approximation is adequate only when the
beam of chiral quasiparticles remains close to the optical axis of the (pseudo-)magnetic lens. In
practice, of course, this condition may break down and, then, one would have to reanalyze
the problem by using numerical methods. In such a regime, various optical aberrations will
appear and further complicate the situation. While all these issues may be of real
importance for making pseudomagnetic lenses in practice, they are beyond the scope
of the conceptual study presented here.

In the case $\mathbf{B}_{\lambda}=0$, the action should describe a free quasiparticle moving
with momentum $\varepsilon/v_F$, i.e.,
\begin{equation}
S_0^{\rm (free)}=\frac{\varepsilon}{v_F}\sqrt{z^2+r_{\perp}^2}\approx \frac{\varepsilon}{v_F} \left( z
+\frac{r_{\perp}^2}{2z} +O(r_{\perp}^4) \right).
\label{lens-eikonal-S0-free}
\end{equation}
In the case of a nonzero $\mathbf{B}_{\lambda}$, we seek $S_0$ in
a similar form, which is given by Eq.~(2) in the main text.
Indeed, by matching the actions at the boundaries of the (pseudo-)magnetic
lens, one can easily show that the eikonal of a quasiparticle moving in (pseudo-)magnetic
fields should also contain only even powers of $r_{\perp}$.

By keeping the terms up to quadratic order in $r_{\perp}^2$ and $B_{\lambda}$, as well as making use of the following relations:
\begin{eqnarray}
\label{lens-eikonal-eqs-be}
\nabla_zS_0 &=& \frac{\varepsilon}{v_F}\left( C +\frac{r_{\perp}^2}{2} A^{\prime}(z) \right),\\
(\bm{\nabla}S_0)^2&=& (\nabla_zS_0)^2 +\left(\frac{\varepsilon r_{\perp}}{v_F}\right)^2(A(z))^2
= \left(\frac{\varepsilon}{v_F}\right)^2C^2 +\left(\frac{\varepsilon r_{\perp}}{v_F}\right)^2\left[C A^{\prime}(z)+(A(z))^2\right],
\label{lens-eikonal-eqs-ee}
\end{eqnarray}
we rewrite Eq.~(\ref{lens-eikonal-epsilon}) as
\begin{eqnarray}
&& C^4 -C^6 +\left(2-3C^2\right)r_{\perp}^2C^2\left[C A^{\prime}(z)+(A(z))^2\right]
+\frac{2\lambda B_{\lambda}}{B^{*}} \left\{C^3 +r_{\perp}^2C\left[\frac{3}{2}C A^{\prime}(z)+(A(z))^2\right]\right\} \nonumber\\
&&
+ \frac{1}{2}\left(\frac{B_{\lambda}}{B^{*}}\right)^2 \left\{C^2 +r_{\perp}^2\left[ CA^{\prime}(z) -2(A(z))^2\right]\right\}
-\frac{r_{\perp}^2 eB_{\lambda}^2}{2c \hbar B^{*}} C^4=0.
\label{lens-eikonal-epsilon-2}
\end{eqnarray}

At the zeroth order in $r_{\perp}^2$, Eq.~(\ref{lens-eikonal-epsilon-2}) reduces to the following
equation:
\begin{eqnarray}
C^4 -C^6
+\frac{2\lambda B_{\lambda}}{B^{*}}  C^3  + \frac{1}{2}\left(\frac{B_{\lambda}}{B^{*}}\right)^2 C^2  &=& 0.
\label{lens-eikonal-epsilon-r-0}
\end{eqnarray}
This equation has four nontrivial solutions, i.e.,
\begin{eqnarray}
C^{(1)}_{\pm} &\simeq& \pm 1 + \lambda \frac{B_{\lambda}}{B^{*}} \mp \frac{5}{4}\left(\frac{B_{\lambda}}{B^{*}}\right)^2
+O\left(\frac{B_{\lambda}^3}{(B^{*})^3}\right),\\
C^{(2)}_{\pm} &\simeq& -\left(\lambda \pm \frac{1}{\sqrt{2}}\right) \frac{B_{\lambda}}{B^{*}} +O\left(\frac{B_{\lambda}^3}{(B^{*})^3}\right),
\label{lens-eikonal-epsilon-r-0-sol}
\end{eqnarray}
where we used an expansion in powers of the small parameter $B_{\lambda}/B^{*}$. By taking into account
that $C$ should be equal to one in the limit of vanishing fields [cf. Eqs.~(2) in the main text and (\ref{lens-eikonal-S0-free})]
we conclude that the physical solution is given by $C=C^{(1)}_{+}$.

Further, by equating the terms quadratic in $r_{\perp}$ in Eq.~(\ref{lens-eikonal-epsilon-2}), we obtain the first-order differential equation (4) in the main text for the function $A(z)$.

\section{Lens equation for the uniform fields}

In this section we present the key steps of the derivation of the lens equation in the simplest
case of uniform magnetic and pseudomagnetic fields, $B_{\lambda}=const$. In the regions $z<0$ and $z>L$, the fields are absent.
Therefore, $a_1=1$ and $a_2=0$ there and the solutions to Eq.~(4) in the main text
are given by
\begin{eqnarray}
A(z)\Big|_{z<0}&=&\frac{1}{z-z_1}, \\
A(z)\Big|_{z>L}&=&\frac{1}{z-z_2},
\label{lens-eikonal-A2-sol-1}
\end{eqnarray}
where $z_1$ and $z_2$ are the integration constants that will be fixed by the boundary conditions.
On the other hand, by solving Eq.~(4) in the main text in the region with nonzero
background fields, i.e., $0<z<L$, we
obtain the following solution:
\begin{equation}
A(z)\Big|_{0<z<L}=a_{2} \cot{\left(\frac{a_{2}}{a_1}z+\phi \right)},
\label{lens-eikonal-A2-sol-2}
\end{equation}
where $\phi $ is another integration constant that should be also determined by matching the solutions
for $A(z)$ at $z=0$ and $z=L$, i.e.,
\begin{eqnarray}
\label{lens-eikonal-A2-matching-be}
-\frac{1}{z_1} &=& a_{2} \cot{\left(\phi \right)},\\
a_{2} \cot{\left(\frac{a_{2}}{a_1}L+\phi \right)} &=& \frac{1}{L-z_2}.
\label{lens-eikonal-A2-matching-ee}
\end{eqnarray}
Finally, after excluding $\phi $ from these equations, we obtain the lens
equation
\begin{equation}
(z_1 + g_{\lambda})(z_2 - h_{\lambda})=-f^2_{\lambda}.
\label{lens-eikonal-lens-eq-main-Supp}
\end{equation}
with
\begin{eqnarray}
\label{lens-eikonal-lens-g-app-2-method}
g_{\lambda} &=& \frac{1}{a_{2}} \cot{\left(\frac{a_{2}L}{a_1}\right)} \simeq \frac{l_{\varepsilon} B^{*}}{B_{\lambda} \left(1-\lambda B_{\lambda}/B^{*}\right)}
\cot{\left(\frac{L B_{\lambda}\left(1-\lambda B_{\lambda}/B^{*}\right)}{l_{\varepsilon} B^{*}}\right)},\\
\label{lens-eikonal-lens-f-app-2-method}
f_{\lambda} &=& \frac{1}{a_{2} \sin{\left(\frac{a_{2}L}{a_1}\right)}} \simeq \frac{l_{\varepsilon} B^{*}}{B_{\lambda} \left(1-\lambda B_{\lambda}/B^{*}\right)
\sin {\left(\frac{L B_{\lambda}\left(1-\lambda B_{\lambda}/B^{*}\right)}{l_{\varepsilon} B^{*}}\right)} },
\end{eqnarray}
and $h_{\lambda} = L+g_{\lambda}$.

\section{Lens equation for the nonuniform fields}

\begin{figure}[t]
\begin{center}
\includegraphics[width=0.5\textwidth]{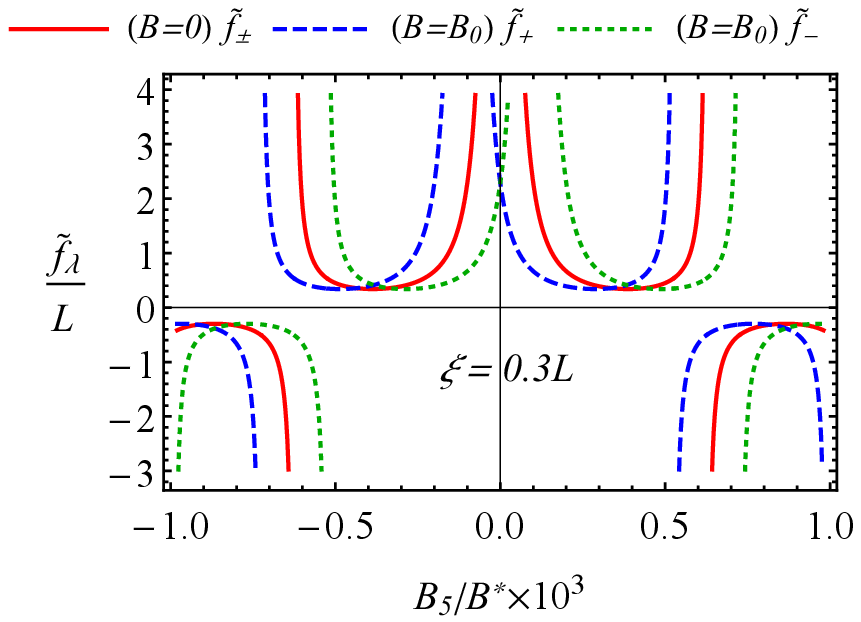}\hfill
\includegraphics[width=0.5\textwidth]{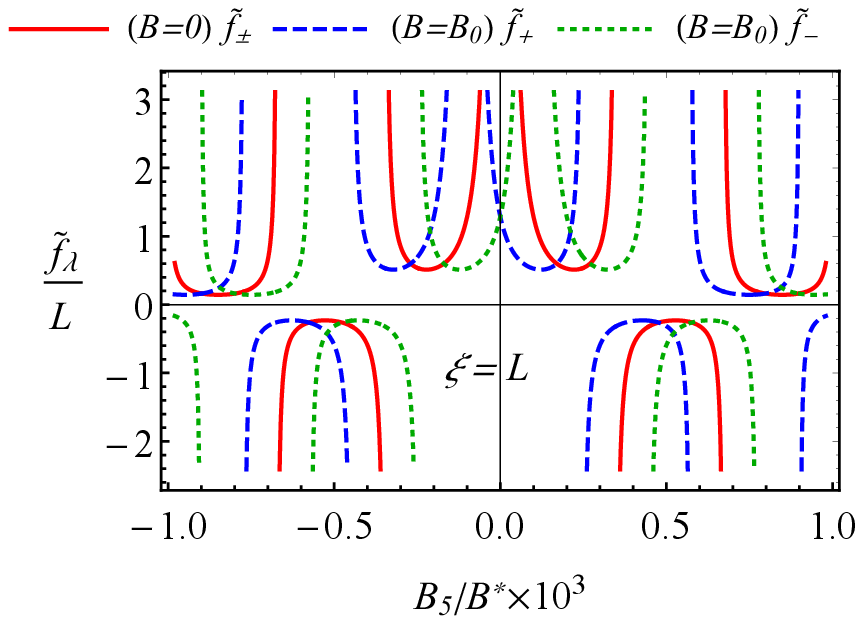}
\end{center}
\caption{The focal length $\tilde{f}_{\lambda}$ for the nonuniform effective field (\ref{inhom-B-profile-Supp}).
While the red solid lines correspond to the quasiparticles of both chiralities at $B=0$, the blue dashed and green dotted lines represent the focal lengths for $\lambda=+$ and $\lambda=-$ at $B_0=10^{-4} B^{*}$, respectively.
The left panel represents the results at $\xi=0.3\,L$ and the right one corresponds to $\xi=L$. We set $\varepsilon=100~\mbox{meV}$ and $L=10^{-2}~\mbox{cm}$.}
\label{fig:focal-inhom-B5-B}
\end{figure}

 In this section we present the derivation of the lens equation as well as focal length
$\tilde{f}_{\lambda}$ for the nonuniform effective magnetic field
\begin{equation}
\tilde{\mathbf{B}}_{\lambda}(z) = \theta(z)\theta\left(L-z\right) \frac{\mathbf{B}_{\lambda}}{1+\left(z-L/2\right)^2/\xi^2},
\label{inhom-B-profile-Supp}
\end{equation}
neglecting the effects of the Berry curvature. In this case, we have $\varepsilon=v_Fp$
instead of Eq.~(1) in the main text. Therefore, the analog of the differential equation
(4) in the main text reads
\begin{equation}
A^{\prime}(z) +(A(z))^2 + \frac{e (\tilde{B}_{\lambda}(z))^2}{2c\hbar B^{*}} =0.
\label{inhom-B-diffeq}
\end{equation}
Taking into account the explicit form of the effective field $\tilde{B}_{\lambda}(z)$ profile (\ref{inhom-B-profile-Supp}), we obtain the following analytical solution inside the solenoid:
\begin{eqnarray}
A(z)\Big|_{0<z<L} =\frac{2c_{0} \xi F_{\lambda} + i c_{0} (L-2z)
-  \left(L-2z+2i\xi F_{\lambda}\right)\exp{\left(2iF_{\lambda}\,
\mbox{arccot}\left(\frac{2\xi}{L-2z}\right)\right)}}{2 \left(\xi^2+\frac{(L-2z)^2}{4}\right)\left[\exp{\left(2iF_{\lambda}\,
\mbox{arccot}\left(\frac{2\xi}{L-2z}\right)\right)}
- ic_{0}\right]},
\label{inhom-B-eikonal}
\end{eqnarray}
where $c_{0}$ is an integration constant and
\begin{equation}
F_{\lambda}\equiv\sqrt{1+\left(\frac{\xi B_{\lambda}}{l_{\varepsilon} B^{*}}\right)^2}.
\label{inhom-lens-F}
\end{equation}
By matching the solutions outside the solenoid (\ref{lens-eikonal-A2-sol-1}) with that in Eq.~(\ref{inhom-B-eikonal})
at $z=0$ and $z=L$, we obtain the following lens equation:
\begin{equation}
(z_1 + \tilde{g}_{\lambda})(z_2 - \tilde{h}_{\lambda})=-\tilde{f}_{\lambda}^2,
\label{inhom-lens-eq}
\end{equation}
where $\tilde{h}_{\lambda} = L+\tilde{g}_{\lambda}$,
\begin{eqnarray}
\tilde{g}_{\lambda} = (L^2+4\xi^2) \frac{2\xi F_{\lambda}  \cos{\left[2F_{\lambda}\, \mbox{arccot}\left(\frac{2\xi}{L}\right)\right]}
+ L  \sin{\left[2F_{\lambda}\, \mbox{arccot}\left(\frac{2\xi}{L}\right)\right]}}{2\left(4\xi^2F_{\lambda}^2-L^2\right)
\sin{\left[2F_{\lambda}\, \mbox{arccot}\left(\frac{2\xi}{L}\right)\right]-8L\xi F_{\lambda} \cos{\left[2F_{\lambda}\,
\mbox{arccot}\left(\frac{2\xi}{L}\right)\right]}} },
\label{inhom-B-g}
\end{eqnarray}
and
\begin{eqnarray}
\tilde{f}_{\lambda} = \frac{\xi \left(L^2+4\xi^2\right)  F_{\lambda}}{\left(4\xi^2F_{\lambda}^2-L^2\right)
\sin{\left[2F_{\lambda}\, \mbox{arccot}\left(\frac{2\xi}{L}\right)\right]}
-4L\xi F_{\lambda} \cos{\left[2F_{\lambda}\, \mbox{arccot}\left(\frac{2\xi}{L}\right)\right]}}.
\label{inhom-B-f}
\end{eqnarray}
Further, it is easy to check that $\lim_{\xi\to\infty}\tilde{f}_{\lambda}=f_{\lambda}^{(0)}$,
where $f_{\lambda}^{(0)}$ is the solution in the case of the homogeneous field
given by Eq.~(10) in the main text. The dependence of the focal length $\tilde{f}_{\lambda} $ on the pseudomagnetic
field strength $B_5$ is presented in Fig.~\ref{fig:focal-inhom-B5-B} for $\xi=0.3\,L$ (left panel) and $\xi=L$ (right panel).
The results for a moderately large $\xi$ depicted in the right panel are almost indistinguishable from
the case of a uniform field, as it should be for a weakly varying (pseudo-)magnetic field.
On the other hand, when the (pseudo-)magnetic field is sufficiently nonuniform, i.e.,
$\xi\lesssim L$, the dependence of $\tilde{f}_{\lambda}$ on the (pseudo-)magnetic field is different.
Comparing the left and right panels in Fig.~\ref{fig:focal-inhom-B5-B}, we see that the period of the focal length oscillations increases with decreasing $\xi$.

%%%%%%%%%%%%%%%%%%
%%%%%%%%%%%%%%%%%%

\vspace{0.5cm}
\begin{center}
\noindent\rule{8cm}{1pt}
\end{center}
\vspace{0.5cm}

\begin{itemize}
\item[[S1\!\!\!]] L.~D.~Landau and E.~M.~Lifshitz, {\it  The Classical theory of fields.} Vol.~2 (Butterworth-Heinemann, Oxford, 1987).
\end{itemize}

\end{document}